\documentclass[a4paper,11pt]{article}
\usepackage{jheppub}
\usepackage{lineno}

\arxivnumber{2610.09216} 

\title{\boldmath Logarithmic corrections to massive particle trajectories for the classical soft log theorems at leading order in $G$}

\author{T.H. Nadai}
\affiliation{University of Campinas — UNICAMP, Institute of Physics “Gleb-Wataghin”,
13083-859, Campinas — SP, Brazil}

\emailAdd{t256856@dac.unicamp.br}

\abstract{We study the logarithmic corrections to the asymptotic trajectories of massive particles. These corrections enter the calculation of the soft-log sector of the radiative field at null infinity in four-dimensional gravitational scattering. We first isolate the part of the metric which enters the particle equations of motion. We show that the nonlinear source can only affect this sector through homogeneous modes on the hyperboloid, and that matching to the proposed soft tower implies that these modes vanish, except for the first mode. We show that this mode cannot affect the trajectory coefficients at leading order in $G$. We then invert the matter energy-momentum tensor on the particle worldlines to obtain the trajectory coefficients at leading order in $G$ and at every logarithmic order using Legendre polynomials. The first two coefficients reproduce the known results at orders $G$ and $G^2$.}

\keywords{Classical Theories of Gravity, Space-Time Symmetries, Scattering Amplitudes}

\begin{document}
\maketitle
\flushbottom

\section{Introduction}
\label{sec:intro}

In this paper we consider a gravitational scattering event in which massive bodies collide. The analysis of the waveform generated by such an event can be extremely complicated. However, if one considers the radiative fields at early and late times, with gravity being the only long range interaction, seemingly universal terms, i.e. terms that do not depend on the details of the scattering, but only on the incoming and outgoing momenta of the particles, appear in the $\mathcal O(r^{-1})$ sector.\footnote{See~\cite{sen2025gravitationalwavetailssoft} for a review of the problem.} These terms have a characteristic $U$ dependence of the form
\begin{equation}
  \label{eq-intro:soft-log-form-retarded-time}
  {(\ln U)^n\over U^{n+1}},\qquad n\geq0
\end{equation}
where $U$ denotes the retarded time corrected to include effects of long range propagation of gravitational waves~\cite{saha2020proofclassicalsoftgraviton}. We refer to these terms as \textit{soft-log} terms, since they first appeared in the context of the soft expansion in four dimensions.

In~\cite{saha2020proofclassicalsoftgraviton,sahoo2020classicalsubsubleadingsoftphoton} the authors have explicitly obtained the coefficients of the $n=0$ and $n=1$ terms by calculating the full energy momentum tensor in momentum space while recently a complementary position space derivation of the $n=0$ term was given in~\cite{compère2026gravitationalmultipolesantipodalmatching} using the Weyl tensor and antipodal matching relations across spatial infinity. The momentum space approach naturally separates into two pieces: (i) the calculation of the matter energy momentum tensor, which involves solving recursively for the trajectories of massive particles, and (ii) the calculation of the non-linear energy momentum tensor. The process starts with a flat background and straight line trajectories. One solves for the first order gravitational perturbation and this first order correction is then substituted into the geodesic equation to obtain the first order correction to the particle's trajectory. For the next iteration, the corrected trajectory is used to calculate the correction to the matter energy momentum tensor and, using the first order gravitational perturbation the non-linear energy momentum tensor is calculated. Combining these contributions one is able to obtain the second order correction to the gravitational field and the process repeats.

In this paper we build upon these works by demonstrating that the matter contribution at leading order in $G$ can be calculated independently of the non-linear contribution, on the condition that there are no terms of the form
\begin{equation}
  \label{eq-intro:forbiden-soft-log-form-retarded-time}
  {(\ln U)^n\over rU^{n}}\quad(n\geq1), \quad {(\ln U)\over r}
\end{equation}
in the classical expansion of the radiative field at null infinity. Under this assumption we calculate an explicit formula for all the leading $G^{n+1}$ trajectory coefficients relevant to the soft-log sector. 

This paper is organized as follows: In section~\ref{sec:setup} we review the setup and our conventions for the problem. In section~\ref{sec:lambda_field} we prove that assuming the absence of terms of the form~\eqref{eq-intro:forbiden-soft-log-form-retarded-time}, the non-linear source does not contribute to the terms necessary to calculate the trajectory coefficients at leading order in $G$. In section~\ref{sec:trajectory_coefficients} we explicitly calculate these coefficients with the general formula~\eqref{eq:exp_for_Cn_simplified}. Our results agree with the results expected from~\cite{saha2020proofclassicalsoftgraviton,sahoo2020classicalsubsubleadingsoftphoton} at orders $G$ and $G^2$. We conclude with a discussion of future directions.

Around the same time as the submission of this work, Karan, Khatun, Mahajan, Sahoo and Sen calculated the all orders classical graviton theorem in the Newtonian limit and conjectured an extension of their expressions to relativistic velocities~\cite{karan2026allordersclassicalsoftgraviton}. Their conjectured expression for the trajectory coefficients in section 5, see equations (5.6) to (5.9), agrees with the expression we obtain at~\eqref{eq:exp_for_Cn_simplified}.

\section{Asymptotic setup}
\label{sec:setup}

\subsection{The \texorpdfstring{$\lambda$}{lambda} order and the trajectory ansatz}

We consider $M\rightarrow N$ scattering. We assume the particles come from past timelike infinity, interact in a finite region for a finite time and then scatter to future timelike infinity. We make no assumptions about what happens \textit{during} the scattering, and we assume that the only long range interaction between the particles in the far past and far future is gravity. Below we shall calculate the \textit{late time} trajectory coefficients, but the early-time analysis proceeds in the same way.

We label the particle's asymptotic momentum by $p_a^\mu$ with $p^2_a = -m_a^2$. We denote the trajectories by $\xi_a(\sigma_a)$ where $\sigma_a$ denotes the proper time of the $a$th particle and the speed of light is set to 1.

To isolate the terms that contribute to the soft-logarithmic tower, we use the $\lambda_\sigma$ order introduced in \cite{karan2025orderclassicalelectromagneticsoft}. This serves solely as a bookkeeping device, not an asymptotic order of magnitude. If a term is of the form
\begin{equation}
  \frac{(\ln\sigma)^a}{\sigma^b},
\end{equation}
then its $\lambda_\sigma$ order is $a-b$. For example, $\ln\sigma/\sigma=\mathcal O(\lambda_\sigma^0)$, $\sigma$ and $\ln\sigma$ both have order $\lambda_\sigma^1$, and $(\ln\sigma)^2/\sigma^5=\mathcal O(\lambda_\sigma^{-3})$. This is not an order of magnitude, since clearly $\sigma\gg(\ln\sigma)^3/\sigma$ as $\sigma\rightarrow\infty$.

We use the same bookkeeping for the large variables $\tau$ and $u$, denoting the corresponding orders by $\lambda_\tau$ and $\lambda_u$.

Following \cite{karan2025orderclassicalelectromagneticsoft}, we take the trajectory ansatz for particle $a$ to be
\begin{equation}
  \label{eq:Trajectory_ansatz}
  \xi_a^\mu(\sigma_a)
  =\frac{p_a^\mu}{m_a}\sigma_a
  +C_0^{(a)\mu}\ln\sigma_a
  -\sum_{n=1}^{\infty}C_n^{(a)\mu}
  \left(\frac{\ln\sigma_a}{\sigma_a}\right)^n
  +\mathcal O(\lambda_\sigma^{-1}),
  \qquad \sigma_a\rightarrow\infty,
\end{equation}
and we seek a general formula for the coefficients $C_n^{(a)\mu}$. A constant offset has been omitted from~\eqref{eq:Trajectory_ansatz}. This omission does not affect our conclusions. We will repeatedly use

\begin{equation}
  \label{eq:def_chi_a}
  \chi_a\equiv\frac{\ln\sigma_a}{\sigma_a}.
\end{equation}

The particle equation of motion is
\begin{equation}
  \label{eq:Equation_of_Motion}
  \ddot\xi_a^\mu(\sigma_a)
  =-\Gamma_{\alpha\beta}^{\mu}[\xi_a(\sigma_a)]
  \dot\xi_a^\alpha\dot\xi_a^\beta.
\end{equation}
Here, the dot denotes differentiation with respect to $\sigma$. In both \cite{saha2020proofclassicalsoftgraviton,sahoo2020classicalsubsubleadingsoftphoton} self-force effects were not relevant to the leading order expressions for $C_0$ and $C_1$ in $G$. Throughout the rest of this work we do not consider self-force contributions.

Plugging~\eqref{eq:Trajectory_ansatz} into the equation of motion gives
\begin{equation}
  \label{eq:trajectory_acceleration}
  \ddot\xi_a^\mu
  =-\frac{1}{\sigma_a^2}
  \left[
    C_0^{(a)\mu}
    +\sum_{n=1}^{\infty}n(n+1)C_n^{(a)\mu}\chi_a^n
  \right]
  +\mathcal O(\lambda_\sigma^{-3}).
\end{equation}
Thus, in order to determine the coefficients $C_n^{(a)}$ we need to compute the right-hand side of~\eqref{eq:Equation_of_Motion} up to order $\lambda_\sigma^{-2}$.

\subsection{The gothic metric and the relevant field sector}

We assume a Minkowski background $\eta=\operatorname{diag}(-1,1,1,1)$ and consider the perturbations
\begin{align}
  -h^{\mu\nu}&=g^{\mu\nu}-\eta^{\mu\nu},
  \\
  \mathfrak h^{\mu\nu}&=\sqrt{-g}\,g^{\mu\nu}-\eta^{\mu\nu}.
\end{align}
The first quantity is the usual perturbation to the metric and the second quantity is the gothic perturbation. We denote $\det g_{\mu\nu} = g$.

The reason we prefer to work with the gothic perturbation is that the harmonic gauge can be easily applied to all orders, simplifying the analysis of higher order terms.

At timelike infinity we use
\begin{equation}
  \tau=\sqrt{-\eta_{\mu\nu}x^\mu x^\nu},
  \qquad
  \rho=\frac{r}{\tau},
  \qquad
  y^\mu=\frac{x^\mu}{\tau}
  =\left(\sqrt{1+\rho^2},\rho\hat x\right).
\end{equation}
The relation between $(\tau,\rho)$ and the retarded time $u=t-r$ is
\begin{equation}
  \label{eq:u_in_timelike}
  u=\tau\left(\sqrt{1+\rho^2}-\rho\right).
\end{equation}
We work under the assumption that the gothic perturbation may be expanded as
\begin{equation}
  \label{eq:Metric_Ansatz}
  \mathfrak h^{\mu\nu}(\tau,\rho,\hat x)
  =\sum_{n=0}^{\infty}\mathfrak h_n^{\mu\nu}(\rho,\hat x)
  \frac{(\ln\tau)^n}{\tau^{n+1}}
  +\mathcal O(\lambda_\tau^{-2}).
\end{equation}
This ansatz is motivated by the equations of motion of point particles. We will see below that if the metric had terms of order $\lambda^{a}_\tau$ with $a\geq0$, then, in general, the acceleration~\eqref{eq:trajectory_acceleration} could have terms of order $\lambda^{-1}_\sigma$ and above that would contradict~\eqref{eq:Trajectory_ansatz}.\footnote{Although such terms could in principle cancel exactly in the geodesic equation, we do not consider this possibility and take~\eqref{eq:Metric_Ansatz} as our starting point in what follows.}

To the order relevant here, the two perturbations are related by
\begin{equation}
  \label{eq:h_in_terms_of_gothic}
  h_{\mu\nu}
  =-\left(
    \mathfrak h_{\mu\nu}
    -\frac12\eta_{\mu\nu}\mathfrak h
  \right)
  +\mathcal O(\lambda_\tau^{-2}),
  \qquad
  \mathfrak h\equiv\eta_{\alpha\beta}\mathfrak h^{\alpha\beta}.
\end{equation}

The Cartesian derivatives in these coordinates are
\begin{align}
  \label{eq:cartesian_partials_in_hyperbolic}
  \partial_t
  &=\sqrt{1+\rho^2}\,\partial_\tau
  -\frac{\rho\sqrt{1+\rho^2}}{\tau}\partial_\rho,
  \\
  \partial_i
  &=-\rho\hat x_i\partial_\tau
  +\frac{1+\rho^2}{\tau}\hat x_i\partial_\rho
  +\frac{1}{\tau\rho}D_i,
\end{align}
where $D_i$ is tangent to the unit sphere. Every Cartesian derivative lowers the $\lambda_\tau$ order in the terms of $\mathfrak h^{\mu\nu}$. It follows from~\eqref{eq:h_in_terms_of_gothic} that $\Gamma_{\alpha\beta}^{\mu}=\mathcal O(\lambda_\tau^{-2})$, and hence only the $\mathcal O(\lambda_\tau^{-1})$ sector of $h_{\mu\nu}$ is required for~\eqref{eq:trajectory_acceleration}. We have used $\lambda_\tau$ and $\lambda_\sigma$ interchangeably since along a late-time particle worldline $\tau=\mathcal O(\sigma_a)$, so calculating $h^{\mu\nu}$ up to $\lambda_\tau^{-1}$ and substituting for $\sigma$ contains all the relevant terms for the trajectory coefficients. This is why a term $\lambda^{0}_\tau$ in~\eqref{eq:Metric_Ansatz} could generate a contribution of $\lambda_\sigma^{-1}$ in~\eqref{eq:trajectory_acceleration}.

In gothic variables the harmonic gauge condition reads
\begin{equation}
  \partial_\mu \mathfrak h^{\mu\nu}=0
\end{equation}
to all orders in $G$. In this gauge the gothic field equation reads
\begin{equation}
  \label{eq:gothic_equation}
  \Box\mathfrak h^{\mu\nu}
  =16\pi G(-g)T^{\mu\nu}+N^{\mu\nu},\qquad \Box\equiv\eta^{\alpha\beta}\partial_\alpha\partial_\beta,
\end{equation}
where
\begin{equation}
  \label{eq:point_particle_stress_tensor}
  T^{\mu\nu}
  =\sum_{b=1}^{N}\int_0^\infty
  m_b\dot\xi_b^\mu(\sigma_b)\dot\xi_b^\nu(\sigma_b)
  \delta^{(4)}(x-\xi_b(\sigma_b))
  \frac{ d\sigma_b}{\sqrt{-g}}.
\end{equation}
We have suppressed the early time trajectory contribution since our analysis will focus on late times after the scattering. The nonlinear source $N^{\mu\nu}$ contains the following structures \cite{blanchet2024postnewtoniantheorygravitationalwaves}.
\[(\mathfrak h)^\ell\partial\mathfrak h \partial \mathfrak h \quad\text{and}\quad (\mathfrak h)^\ell\mathfrak h \partial^2\mathfrak h,\]
with $\ell\geq0$. We will see that the explicit form of the non-linear source will not be relevant. We write
\begin{equation}
  \label{eq:matter_plus_Q}
  \mathfrak h^{\mu\nu}
  =\mathfrak h_M^{\mu\nu}+Q^{\mu\nu},
  \qquad
  \mathfrak h_M^{\mu\nu}
  =16\pi G\Box^{-1}[(-g)T^{\mu\nu}],
  \qquad
  Q^{\mu\nu}=\Box^{-1}[N^{\mu\nu}].
\end{equation}
Here $\Box^{-1}$ is the retarded inverse.
We now determine whether $Q^{\mu\nu}$ can contribute at order $\lambda_\tau^{-1}$.

\section{The \texorpdfstring{$Q^{\mu\nu}$}{Qmunu} gravitational contribution}
\label{sec:lambda_field}
\subsection{The retarded matter field at large \texorpdfstring{$\rho$}{rho}}

We start by calculating the general form of the matter contribution at large $\rho$. The large-$\rho$ behavior gives the basis for the induction argument below. Note that performing a large $\tau$, large $\rho$ limit is equivalent to taking the null infinity limit. However, we keep working in $(\rho,\tau)$ variables since we have more control over the behavior of $\ln\rho$ and $\ln\tau$ than over $\ln r$. By large $(\tau,\rho)$, we mean the limit where:
\begin{equation}
  \tau\gg\rho\gg1
\end{equation}
which has an overlap with the region described by:
\begin{equation}
  r\gg u\gg1.
\end{equation}
In particular, for the second region we first take $r\rightarrow\infty$ at fixed $u$ and then take $u\rightarrow\infty$. In this region, we can make matching arguments with the previous results involving the classical soft graviton theorem.

With the retarded Green function, the contribution from the point particles at late times is
\begin{equation}
  \label{eq:retarded_matter_field}
  \mathfrak h_M^{\mu\nu}(x)
  =-16\pi G\sum_{b=1}^N\int_0^\infty
  \frac{m_b}{2\pi}
  \dot\xi_b^\mu(\sigma_b)\dot\xi_b^\nu(\sigma_b)
  \delta_+\bigl(-(x-\xi_b(\sigma_b))^2\bigr)
  \sqrt{-g}\,d\sigma_b.
\end{equation}
Here
\begin{equation}
  \delta_+\bigl(-(x-\xi_b(\sigma_b))^2\bigr)
  \equiv
  \delta\bigl(-(x-\xi_b(\sigma_b))^2\bigr)
  \Theta(x^0-\xi_b^0(\sigma_b))
\end{equation}
and $\Theta$ denotes the Heaviside step function. This means we must evaluate the delta condition and select the branch in the past light cone. Performing the $\sigma_b$ integral in~\eqref{eq:retarded_matter_field} gives
\begin{equation}
  \label{eq:matter_field_on_root}
  \mathfrak h_M^{\mu\nu}(x)
  =-4G\sum_{b=1}^N m_b
  \left.
  \frac{\dot\xi_b^\mu\dot\xi_b^\nu}
  {|(x-\xi_b)\cdot\dot\xi_b|}
  \sqrt{-g}
  \right|_{\sigma_b=\sigma_b(\tau,y)}.
\end{equation}

We have suppressed the argument of $\xi_b$.
For large $\rho$, let $q^\mu=(1,\hat x)$. The denominator behaves as
\begin{equation}
  \label{eq:inverting_T_mu_nu}
  |(x-\xi_b)\cdot\dot\xi_b|
  =\tau\rho|q\cdot\dot\xi_b|
  \left[1+\mathcal O\!\left((\tau\rho)^{-1}\right)\right],
\end{equation}
where we used $x^\mu\simeq\tau\rho q^\mu$ at large $\rho$. Thus the leading part is precisely the matter $r^{-1}$ field at null infinity. We assume that the determinant on a worldline admits the expansion
\begin{equation}
  \label{eq:metric_det_exp}
  \sqrt{-g(\xi_b(\sigma_b))}
  =1+\sum_{n=1}^{\infty}\sum_{k=0}^{n-1}
  B_{kn}^{(b)}\frac{(\ln\sigma_b)^k}{\sigma_b^n}.
\end{equation}
This uses~\eqref{eq:Metric_Ansatz} together with
\[\tau(\sigma)=\sqrt{-\xi^2}=\mathcal O(\lambda_\sigma),\qquad \rho(\sigma)=\frac{|\vec \xi|}{\sqrt{-\xi^2}}=\mathcal O(\lambda_\sigma^0).\]

After evaluating the delta condition in~\eqref{eq:retarded_matter_field} with the trajectory~\eqref{eq:Trajectory_ansatz} we obtain:
\begin{equation}
  \sigma_b(\tau,y)\sim u\left(1+\mathcal O(\lambda_u^0)\right)+\mathcal O(r^{-1}).
\end{equation}

From~\eqref{eq:u_in_timelike}, at large $u$ we get $u\sim \tau/(2\rho)$, which allows us to expand~\eqref{eq:matter_field_on_root} as
\begin{equation}
  \label{eq:hmatterforlargerho}
    \mathfrak h_M\sim {4G\over\rho\tau}
    A_{00}+{4G}\left(\sum_{n=1}^{\infty}\frac{\rho^{n-1}}{\tau^{n+1}}\sum_{k+\ell\leq n-1}A_{\ell kn}{(\ln\rho)^\ell(\ln\tau)^k}\right)+\mathcal O(r^{-2}).
\end{equation}
Thus the part of the matter field which behaves as $\rho^{n-1}/\tau^{n+1}$ obeys
\begin{equation}
  \label{eq:matter_bound}
  \left.
  \mathfrak h_M
  \right|_{\rho^{n-1}\tau^{-n-1}}
  =\mathcal O\!\left(
    \frac{\rho^{n-1}}{\tau^{n+1}}
    (\ln\rho)^\ell(\ln\tau)^k
  \right),
  \qquad
  0 \leq k+\ell\leq n-1,
  \qquad n\geq1.
\end{equation}
We see that the total logarithmic degree is at most $n-1$. This behavior was expected since it reflects the soft $\lambda^{-1}_u$ tower coming from matter terms.

We point out that~\eqref{eq:hmatterforlargerho} has only one term at $\lambda_\tau^{-1}$ while all subleading terms are at most $\lambda_\tau^{-2}$. This does not mean that $\mathfrak h_M$ has no contributions at the $\lambda_\tau^{-1}$ sector. Indeed,
\begin{equation}
  \frac{\rho^{n-1}}{\tau^{n+1}}
  =\frac{1}{2^n r u^n}+\mathcal O(r^{-2})
\end{equation}
while the $\lambda^{-1}_\tau$ sector from $\mathfrak h_M$ behaves at large $\rho$ as $1/(\rho\tau)^{n+1}$ and therefore begins at order $r^{-n-1}$, rather than $1/(ru^n)$. They are precisely the terms that will be used later to determine the trajectories, but become suppressed in a null infinity limit. Here we use the regular $u=t-r$ retarded time instead of the capital $U$ for the gravitationally corrected retarded time from~\cite{saha2020proofclassicalsoftgraviton}:

\begin{equation}
  \label{eq-Qmunu:corrected-retarded-time}
  U = u+2G\left(\sum_b p_b \cdot q\right)\ln r.
\end{equation}

\subsection{Homogeneous blocks}

We suppress tensor indices in this subsection. We write the polyhomogeneous expansion of the nonlinear field as
\begin{equation}
  \label{eq:Q_full_expansion}
  Q
  =\sum_{a=1}^{\infty}\sum_{k=0}^{\infty}
  Q_k^a(\rho,\hat x)
  \frac{(\ln\tau)^k}{\tau^{k+a}}, \qquad \Box Q=N.
\end{equation}
Here $Q^{a}_k$ belongs to order $\lambda^{-a}_\tau$. Since $N$ begins at $\mathcal O(\lambda_\tau^{-4})$, the $\lambda^{-1}_\tau$ coefficients satisfy
\begin{equation}
  \label{eq:harmonic_Q}
  \mathcal H_n Q_n^1=0,
  \qquad
  \mathcal H_n\equiv\Delta_{H^3}-(n^2-1).
\end{equation}
Here $\Delta_{H^3}$ is the Laplacian on the unit hyperboloid. We show that $Q_n^1$ will not contribute to the trajectory coefficients at leading order. Note that we cannot discard it as homogeneous data, since a homogeneous mode on the hyperboloid does not need to be excluded by the no-incoming-radiation prescription. Our strategy will be to divide the coupled equations into blocks with the same denominator.

Terms with the same denominator form a closed block,
\begin{equation}
  \label{eq:def_Q_block}
  \mathcal Q_n
  \equiv\frac{1}{\tau^{n+1}}
  \sum_{a=1}^{n+1}
  Q_{n+1-a}^{a}(\ln\tau)^{n+1-a}.
\end{equation}
Indeed,
\begin{align}
  \label{eq:box_on_Q_block_term}
  \Box\left(
    Q_{n+1-a}^{a}
    \frac{(\ln\tau)^{n+1-a}}{\tau^{n+1}}
  \right)
  =\frac{1}{\tau^{n+3}}
  \Bigl[&
    (\ln\tau)^{n+1-a}\mathcal H_n Q_{n+1-a}^{a}
    \nonumber\\
    &+2n(n+1-a)(\ln\tau)^{n-a}Q_{n+1-a}^{a}
    \nonumber\\
    &-(n+1-a)(n-a)(\ln\tau)^{n-1-a}Q_{n+1-a}^{a}
  \Bigr].
\end{align}
For instance the equations for a block with denominator $\tau^{n+1}$ are:\footnote{When replacing $(n,a)$ values, $Q^b_m$ with $b<1$ should be understood as 0.}
\begin{align}
  \nonumber
  &\mathcal H_n Q_n^1=0
  \\
  &
  \nonumber
  \mathcal H_n Q_{n-1}^2
    +2n^2Q_n^1 = N\left[\frac{(\ln\tau)^{n-1}}{\tau^{n+3}}\right]
  \\
  &\vdots
  \notag\\
  &
  \label{eq:equation_blocks}
  \mathcal H_n Q_{n+1-a}^{a}
    +2n(n+2-a)Q_{n+2-a}^{a-1}
    -(n+3-a)(n+2-a)Q_{n+3-a}^{a-2}\\
    &
    \nonumber=N\left[\frac{(\ln\tau)^{1+n-a}}{\tau^{n+3}}\right],
    \qquad 3\leq a\leq n
  \\
  &\vdots
  \notag\\
  &
  \nonumber
  \mathcal H_n Q_0^{n+1}
    +2nQ_1^n
    -2Q_2^{n-1}=N\left[\frac{1}{\tau^{n+3}}\right].
\end{align}
Here $N[f(\tau)]$ indicates the coefficients of $f(\tau)$ in the non-linear source expansion.
Note that the block for $n=0$ contains only $Q_0^1$. We will return to this block at the end. For $a>1$, we separate each coefficient as
\begin{equation}
  Q_{n+1-a}^{a}=S_{n+1-a}^{a}+\varpi_{n+1-a}^{a},
\end{equation}
where $S$ is obtained by inverting $\mathcal H_n$ on each equation in~\eqref{eq:equation_blocks}, while any homogeneous term which could be present is absorbed into $\varpi$. The special case $a=1$ has $S_n^1=0$ and $Q_n^1=\varpi_n^1$. Thus we have broken $\mathcal Q_n$ into one homogeneous block $\Pi_n$ and one sourced block $\mathcal S_n$:
\begin{equation}
  \mathcal Q_n=\mathcal S_n+\Pi_n,
  \qquad
  \Box\mathcal S_n=\mathcal N_n,
  \qquad
  \Box\Pi_n=0.
\end{equation}
Here $\mathcal N_n$ denotes the part of the nonlinear source with denominator $\tau^{n+3}$.
All the terms related to $Q_n^1$ have been moved to $\Pi_n$. Although $\Pi_n$ satisfies a homogeneous equation for the d'Alembertian, it cannot be set to zero immediately, since we are not directly inverting the source with the Green function integral. This essentially means that the homogeneous block captures the residual ambiguity after inverting solely with $\mathcal H_n^{-1}$ in the timelike region.

First, let $n>0$ and
\begin{equation}
  \kappa=-y\cdot q,
  \qquad
  q^\mu=(1,\hat q),
  \qquad
  q^2=0.
\end{equation}
Here $q^\mu$ now denotes the null direction integrated over the boundary sphere.
We may solve the coupled homogeneous equations in terms of bulk-to-boundary propagators~\cite{Campiglia_2015}. The first equation introduces an arbitrary boundary coefficient $A_n^1(\hat q)$,
\begin{equation}
  \label{eq:first_term_in_the_tower}
  Q_n^1
  =\int d^2\hat q\,
  \frac{A_n^1(\hat q)}{\kappa^{n+1}}.
\end{equation}
The second equation determines the part of $\varpi_{n-1}^2$ generated by $A_n^1$, but it also introduces a new homogeneous coefficient $A_{n-1}^2(\hat q)$:
\begin{equation}
  \varpi_{n-1}^2
  =\int d^2\hat q\,
  \frac{nA_n^1(\hat q)\ln\kappa+A_{n-1}^2(\hat q)}
  {\kappa^{n+1}}.
\end{equation}
Continuing in this way, every lower equation introduces one new arbitrary boundary coefficient. The general coefficient is
\begin{equation}
  \label{eq:general_varpi_coefficient}
  \varpi_{n+1-a}^{a}
  =\int d^2\hat q\,
  \frac{1}{\kappa^{n+1}}
  \sum_{j=n+1-a}^{n}
  \binom{j}{n+1-a}
  A_j^{n+1-j}(\hat q)(\ln\kappa)^{j-(n+1-a)},
  \qquad 1\leq a\leq n+1.
\end{equation}
Joining the coefficients with their corresponding powers of $\ln\tau$ gives
\begin{align}
  &\Pi_n=\frac{1}{\tau^{n+1}}
  \left[
    Q_n^1(\ln\tau)^n
    +\varpi_{n-1}^2(\ln\tau)^{n-1}
    +\varpi_{n-2}^3(\ln\tau)^{n-2}
    +\cdots
    +\varpi_0^{n+1}
  \right]
  \nonumber\\
  &=
  \frac{1}{\tau^{n+1}}
  \int d^2\hat q\,
  \frac{1}{\kappa^{n+1}}
  \sum_{b=0}^{n}
  \sum_{j=b}^{n}
  \binom{j}{b}
  A_j^{n+1-j}(\hat q)
  (\ln\tau)^b
  (\ln\kappa)^{j-b}.
\end{align}
Using the binomial theorem, the complete solution containing all the independent chains becomes
\begin{equation}
  \label{eq:complete_Pi_block}
  \Pi_n(\tau,y)
  =\int d^2\hat q\,
  \frac{1}{(\tau\kappa)^{n+1}}
  \sum_{j=0}^{n}
  A_j^{n+1-j}(\hat q)[\ln(\tau\kappa)]^j.
\end{equation}
Since $\tau\kappa=-x\cdot q$ and $q^2=0$, every term in~\eqref{eq:complete_Pi_block} separately solves the homogeneous wave equation.

We now evaluate its leading logarithm at future null infinity. For large $\rho$~\cite{Campiglia_2015},
\begin{align}
  &\kappa
  =\sqrt{1+\rho^2}-\rho\hat x\cdot\hat q
  =\rho(1-\hat x\cdot\hat q)+\frac{1}{2\rho}
  +\mathcal O(\rho^{-3}),\\
  & \label{eq:bulk_to_boundary_delta}
  \frac{1}{\kappa^{n+1}}
  \longrightarrow
  \frac{2^{n+1}\pi}{n}\rho^{n-1}
  \delta^{(2)}(\hat x,\hat q)
  +\mathcal O(\rho^{n-3}),\quad n\geq 1
\end{align}
the solution for $\Pi_n$ has the leading behavior
\begin{equation}
  \label{eq:forbidden_u}
  \Pi_n(\tau,\rho,\hat x)
  =\frac{2\pi}{n}A_n^1(\hat x)
  \frac{(\ln u)^n}{ru^n}
  +\mathcal O\!\left(
    \frac{(\ln u)^{n-1}}{ru^n}
  \right).
\end{equation}
The proposed soft hierarchy at this order starts at $(\ln u)^{n-1}/(ru^n)$ \cite{saha2020proofclassicalsoftgraviton}.\footnote{We can directly compare the $1/r$ sector of the gothic perturbation $\mathfrak h$ with the regular perturbation $h$ since we can also derive the relation~\eqref{eq:h_in_terms_of_gothic} for the expansion in $r$.} The full solution, however, is given by $\mathcal Q_n=\Pi_n + \mathcal S_n$; thus this term need not vanish if it cancels with a contribution of the sourced block. In the next subsection we will show that the sourced block never reaches the $\lambda_u^{0}$ sector at $r^{-1}$. We remark that the soft hierarchy sets a bound for $U$, however, the presence of a $\lambda_u^0$ term would also imply the presence of a $\lambda_U^0$ term, by~\eqref{eq-Qmunu:corrected-retarded-time}.

For the special case of $n=0$, we have
\begin{equation}
  \Box\left({Q_0^1(\rho,\hat x)\over\tau}\right)=0
\end{equation}
and the asymptotic behavior of this solution is given by
\begin{equation}
  {Q_0^1(\rho,\hat x)\over\tau} \sim A_0^1(\hat x)\frac{\ln\rho}{\tau\rho}+\frac{B_0^1(\hat x)}{\tau\rho}.
\end{equation}

The branch with $\ln\rho$ would generate terms that behave as $(\ln u)/r$. Under the same hypothesis above, we conclude that this term must be zero since, by definition of $\mathcal S_n$, there is no term in the sourced block that can cancel this contribution. The $B_0^1$ branch is allowed and must be at least $G^2$, since it comes from the non-linear source.

\subsection{The nonlinear source bound}

We now show inductively that the nonlinear source cannot contain the term required to cancel $A_n^1(\hat x)$. We first work through the $n=1$ and $n=2$ sectors explicitly, and then use the same bound for a general block.

We have placed every homogeneous solution in $\Pi_n$, so the equations for the sourced coefficients are inverted with a fixed $\mathcal H_n^{-1}$. Since we are interested in seeing whether the sourced block can restore the soft hierarchy \textit{at null infinity}, it suffices to examine the behavior of the non-linear source at large $\rho$ and $\tau$.\footnote{The Green function $G_n(y,y')$ for $\mathcal H_n^{-1}$ at finite $\rho'\ll\rho$ behaves as $\rho^{-n-1}f(y',\hat q)$. Thus, a source contribution
from any bounded region of $\rho'$ has this decaying behavior. The growing
$\rho^{n-1}$ behavior relevant here can arise only from the large-$\rho'$
tail.}

The non-linear source could preserve the soft-hierarchy without requiring $A_n^{1}=0$ if a particular $S^a_{n+1-a}$, combined with its $\tau$-dependent factor, contains enough powers of $\ln\rho$ to produce a term of the form
\begin{equation}
  \label{eq:dangerous_source_after_inversion}
  \frac{\rho^{n-1}}{\tau^{n+1}}(\ln\tau)^{n+1-a}(\ln\rho)^{a-1}\rightarrow \frac{1}{r}\left(\frac{\ln u}{u}\right)^n + \dots
\end{equation}

Usually, applying $\mathcal H_n^{-1}$ does not raise logarithmic powers. However, in the special case where $\rho$ is raised to one of the indicial roots of $\mathcal H_n$, the inverse can add one logarithm,\footnote{The special case $n=0$ can add two logarithms, but $\mathcal H_0$ is never used to invert a source term.}
\begin{equation}
  \label{eq:Hn_raises_log}
  \mathcal H_n^{-1}
  \left[\rho^{n-1}(\ln\rho)^b\right]
  \sim\rho^{n-1}(\ln\rho)^{b+1}.
\end{equation}
Thus, in the expression of the non-linear source for the coefficient $S_{n+1-a}^a$, a term capable of generating the unwanted contribution at $\lambda_u^0$ would behave as:
\begin{equation}
  \label{eq:dangerous_source_term}
  \frac{\rho^{n-1}}{\tau^{n+3}}
  (\ln\tau)^{n+1-a}(\ln\rho)^{a-2}
  \qquad 2\leq a\leq n+1
\end{equation}
at large $\rho$. We want to show that no such term is present.

For $n=1$, the sourced block is simply
\begin{equation}
  \mathcal S_1=\frac{S_0^2}{\tau^2},
  \qquad
  \mathcal H_1S_0^2
  =N\!\left[\frac{1}{\tau^4}\right].
\end{equation}
Recall there is no $a=1$ term for the sourced block since $N^{\mu\nu}$ starts at $\lambda^{-4}_\tau$. At this order the source is built solely from the $\tau^{-1}$ terms in $\mathfrak h$: the leading matter field $\mathfrak h_M|_{\tau^{-1}}$ and $Q_0^1/\tau$, both of which behave as $(\tau\rho)^{-1}$. Therefore
\begin{equation}
  N\!\left[\frac{1}{\tau^4}\right]
  \sim\frac{1}{(\tau\rho)^4},
\end{equation}
and $\mathcal H_1^{-1}$ cannot produce a term in $S_0^2$ which behaves as $\ln\rho$. Thus we conclude:
\begin{align}
  &Q_1^{1}=0,
  &Q_0^{2}&\quad\text{has no term proportional to $\ln\rho$}.
\end{align}

This conclusion holds at every order in $G$. Higher orders in $G$ may change the coefficients of $\mathfrak h$, but they do not change the corresponding $(\tau,\rho)$ asymptotic dependence.
We next consider $n=2$. The sourced equations are
\begin{align}
  \mathcal H_2S_1^2
  &=N\!\left[\frac{\ln\tau}{\tau^5}\right],
  \\
  \mathcal H_2S_0^3+4S_1^2
  &=N\!\left[\frac{1}{\tau^5}\right].
\end{align}
The first source is built from the $\lambda_\tau^{-1}$ matter field, so $S_1^2$ has no term that behaves as $\rho\ln\rho$. For the second equation, the possible couplings at order $\tau^{-5}$ are
\begin{multline}
  N\{\mathfrak h|_{\tau^{-1}},\mathfrak h|_{\tau^{-1}},
  \mathfrak h|_{\tau^{-1}}\},
  \qquad
  N\{\mathfrak h|_{\tau^{-1}},
  \mathfrak h_M|_{(\ln\tau)\tau^{-2}}\},
  \\
  N\left\{\mathfrak h|_{\tau^{-1}},{Q_0^2\over\tau^2}\right\},
  \qquad
  N\{\mathfrak h|_{\tau^{-1}},
  \mathfrak h_M|_{\tau^{-2}}\}.
\end{multline}
The first two couplings involve only the $\lambda_\tau^{-1}$ sector and cannot generate $\rho(\ln\rho)^2$ after inversion. The third does not contribute to the required logarithmic structure because the $n=1$ analysis has already shown that $Q_0^2$ contains no $\ln\rho$ term. We are left with the last coupling. By~\eqref{eq:matter_bound},
\begin{equation}
  \mathfrak h_M|_{\tau^{-2}}=\mathcal O(\tau^{-2}\rho^0),
  \qquad
  \partial\mathfrak h_M|_{\tau^{-2}}
  =\mathcal O(\tau^{-3}\rho),\qquad \partial^2\mathfrak h_M|_{\tau^{-2}}
  =\mathcal O(\tau^{-4}\rho^2)
\end{equation}
whereas $\mathfrak h|_{\tau^{-1}}\sim(\tau\rho)^{-1}$ and its derivative is of order $(\tau\rho)^{-2}$. Thus
\begin{align*}
  &N\{\mathfrak h|_{\tau^{-1}},
  \mathfrak h_M|_{\tau^{-2}}\}
  =\mathcal O\!\left(\frac{\rho}{\tau^5}\right),\\
  &\implies S_0^3
  =\mathcal H_2^{-1}\left(
  N\!\left[\frac{1}{\tau^5}\right]
  -4S_1^2\right)
  \quad\text{has no term proportional to $\rho(\ln\rho)^2$}.
\end{align*}
Therefore $Q_2^1$ must also vanish. What we have essentially shown in the first two cases is that, for every term, the radial and logarithmic behavior satisfies
\begin{equation}
  \label{eq:metric_induction_bound}
  \mathfrak h\big|_{\tau^{-n-1}}
  =\mathcal O\!\left(
  \frac{\rho^{n-1}}{\tau^{n+1}}
  (\ln\tau)^{n+1-a}(\ln\rho)^{a-2}
  \right),
  \qquad n+1\geq a>1,
\end{equation}
at every order in $G$. The highest power of $\rho$ is two below the power in the $\tau$ denominator and, for those terms, the total logarithmic degree is also at most two below it.

We now assume that this holds up to $n=p-1$ and show that it also holds for $n=p$. The nonlinear source contains at least two metric factors and two derivatives. A Cartesian derivative lowers the $\tau$ power while it may raise the $\rho$ power, so the difference between the $\rho$ and $\tau$ exponents cannot increase. At fixed denominator $\tau^{p+3}$, a term containing $\ell$ metric factors can behave at most as $\rho^{p+3-2\ell}$. We thus see that only the quadratic terms can reach $\rho^{p-1}$.

We now look at the logarithms. Before differentiation, the two $\tau$ denominators must add to $p+1$. This can happen either by combining the leading $\tau^{-1}$ field with a $\tau^{-p}$ field, or by combining two fields whose denominators are both larger than one. In the first case, the leading field carries no logarithm and the other factor obeys~\eqref{eq:metric_induction_bound}. After the two derivatives, the source has total logarithmic degree at most $p-2$. In the second case, both factors obey the subleading bound and the total logarithmic degree is at most $p-3$. Neither case can produce
\begin{equation}
  \frac{\rho^{p-1}}{\tau^{p+3}}
  (\ln\tau)^{p+1-a}(\ln\rho)^{a-2},
\end{equation}
whose total logarithmic degree is $p-1$. Inverting $\mathcal H_p$ may add one $\rho$ logarithm, but it still cannot generate the homogeneous chain required to cancel $A_p^{1}(\hat x)$. The soft hierarchy therefore sets $Q_p^1= 0$, for $p>0$, and preserves the bound~\eqref{eq:metric_induction_bound} through $n=p$. Since there is no possible quantity that can cancel~\eqref{eq:forbidden_u} we conclude that
\begin{equation}
  \label{eq:conclusion_of_the_arg}
  \mathfrak h|_{\lambda_\tau^{-1}} = \mathfrak h_M|_{\lambda^{-1}_\tau}+\frac{Q_0^1}{\tau}.
\end{equation}

The Christoffel symbol is linear at leading $\lambda_\tau$ order:
$$\Gamma\left[\mathfrak h|_{\lambda^{-1}_\tau}\right] = \Gamma\left[\mathfrak h_M|_{\lambda^{-1}_\tau}\right]+\Gamma\left[{Q_0^1\over\tau}\right]+\mathcal O(\lambda_\tau^{-3})$$
However, $Q$ begins at order $G^2$. Therefore, at leading order in $G$, $C_0$ is given solely in terms of $\mathfrak h_M$. As we shall see, each subsequent coefficient $C_n$ calculated from $\mathfrak h_M$ is built from $C_0$ and shows up at order $G^{n+1}$. Since
\begin{equation*}
  \frac{\xi^\mu}{\sigma} = \frac{p^\mu}{m} + C_0^\mu\frac{\ln\sigma}{\sigma} + \mathcal O(\lambda_\sigma^{-1})
\end{equation*}
every correction which might come from $Q$ in the equations of motion, by expanding in $\sigma$, would come attached to a factor of $G$ from $C_0$. Moreover, the $G^2$ correction to $C_0$ would always supply one additional power of $G$ for each coefficient. Thus for each $n$, $Q$ only corrects $C_n$ at order $G^{n+2}$. Therefore, the leading part of each trajectory coefficient can be calculated using only $\mathfrak h_M$.

\section{Trajectory coefficients}
\label{sec:trajectory_coefficients}

After showing~\eqref{eq:conclusion_of_the_arg} the derivation closely follows that of~\cite{karan2025orderclassicalelectromagneticsoft}. We calculate~\eqref{eq:retarded_matter_field} for $x=\xi_a(\sigma_a)$ and we sum over all outgoing particles $b\neq a$. The full determinant will supply corrections of order $\lambda_\sigma^{-2}$ to $\mathfrak h_M$ so it can be set to 1 at $\mathcal O(\lambda_\sigma^{-1})$.

After evaluating the delta condition we obtain the expression for the proper time root of particle $b$ in terms of the proper time of particle $a$:\footnote{At this order we may replace $\ln\sigma_b\rightarrow\ln\sigma_a$ since their difference is $\mathcal O(\lambda_\sigma^0)$.}

\begin{align}
  \nonumber
  \sigma_b
  =&
  \nonumber
  -\frac{p_a\cdot p_b}{m_a m_b}\sigma_a
  +\frac{p_b\cdot\Delta C_{ba}}{m_b}\ln\sigma_a
  \\
  \nonumber
  &-\sigma_a\Bigg[\left(\left(\frac{p_a\cdot p_b}{m_a m_b}\right)^2-1\right)-2\chi_a\Bigg(\frac{(p_a\cdot p_b)(p_b\cdot\Delta C_{ba})}
  {m_a m_b^2}+\frac{p_a\cdot\Delta C_{ba}}{m_a}\Bigg)\\
  &
  \label{eq:sigma_b_in_terms_of_sigma_a}
  \quad+\chi_a^2 \Bigg(\left(\frac{p_b\cdot\Delta C_{ba}}{m_b}\right)^2
  +\Delta C_{ba}^2\Bigg)\Bigg]^{1\over2}+\mathcal O(\lambda_\sigma^{0})
\end{align}
where
\begin{equation}
  \label{eq:definition_Delta_C}
  \Delta C_{ba}^\mu = C^{(b)\mu}_0 - C^{(a)\mu}_0
\end{equation}
and $\chi_a$ was defined in~\eqref{eq:def_chi_a}. Before moving on it is convenient to define the following quantities:
\begin{align}
  \label{eq:A_definition}
  A_{ba}^\mu
  &\equiv\frac{p_a^\mu}{m_a}
  +\frac{p_b^\mu}{m_b^2}\frac{p_a\cdot p_b}{m_a},
  \\
  \label{eq:B_definition}
  B_{ba}^\mu
  &\equiv\Delta C_{ba}^\mu
  +\frac{p_b^\mu}{m_b^2}(p_b\cdot\Delta C_{ba}),\\
  \label{eq:def_mathcal_P}
  \mathcal P_{b,\mu\nu}
  &\equiv p_{b,\mu} p_{b,\nu}+\frac12\eta_{\mu\nu}m_b^2.
\end{align}
With these definitions, the square root in~\eqref{eq:sigma_b_in_terms_of_sigma_a} becomes
\begin{equation}
  \label{eq:root_with_A_and_B}
  \sqrt{A_{ba}^2-2\chi_a (A_{ba}\cdot B_{ba})+B_{ba}^2\chi_a^2}
\end{equation}
where contractions are made with the Minkowski inner product. After simplifying, we obtain the derivative of the usual perturbation through order $\lambda_\sigma^{-2}$, summed over outgoing particles $b\neq a$:

\begin{equation}
  \label{eq:usual_pert_derivative}
  \partial_\alpha h_{\mu\nu}|_{x=\xi_a(\sigma_a)}=\sum_{b\neq a}
  \frac{4G}{m_b\sigma_a^2}\frac{B_{ba,\alpha}\chi_a-A_{ba,\alpha}}{\left(A_{ba}^2-2\chi_a (A_{ba}\cdot B_{ba})+B_{ba}^2\chi_a^2\right)^{3\over2}}\mathcal P_{b,\mu\nu}+\mathcal O(\lambda_\sigma^{-3}).
\end{equation}
The fraction in the middle may be expanded in terms of Legendre polynomials using
\begin{align}
  \label{eq:legendre_expansion_for_derivative}
  \frac{B_{ba,\alpha}\chi_a-A_{ba,\alpha}}
  {\left(A_{ba}^2-2\chi_a (A_{ba}\cdot B_{ba})+B_{ba}^2\chi_a^2\right)^{3\over2}}
  =\sum_{n=0}^{\infty}
  &\frac{1}{(A^2_{ba})^{3/2}}
  \left(\frac{B_{ba}^2}{A^2_{ba}}\right)^{n/2}
  \chi_a^n
  \nonumber\\
  &\times\Bigg[
  B_{ba,\alpha}\sqrt{\frac{A_{ba}^2}{B_{ba}^2}}\,P_n'(v_{ba})
  -A_{ba,\alpha}P_{n+1}'(v_{ba})
  \Bigg].
\end{align}
Here the prime denotes differentiation with respect to the argument, and $v_{ba}$ is defined by:
\begin{equation}
  \label{eq:def_of_v}
  v_{ba}\equiv\frac{A_{ba}\cdot B_{ba}}{\sqrt{A_{ba}^2B_{ba}^2}}.
\end{equation}

Using~\eqref{eq:usual_pert_derivative} to compute the Christoffel symbols, we substitute them into~\eqref{eq:Equation_of_Motion} and obtain:
\begin{align*}
\ddot\xi_a^\mu
&=
\begin{aligned}[t]
-\frac12
\eta^{\mu\nu}\frac{1}{\sigma_a^2}
\sum_{n=0}^{\infty}
\left(\frac{\ln\sigma_a}{\sigma_a}\right)^n
\sum_{b\neq a}^N
\frac{4G}{m_a^2m_b}
\frac{1}{(A_{ba}^2)^{3/2}}
\left(\frac{B_{ba}^2}{A_{ba}^2}\right)^{n/2}
\Bigg[
P_n'(v_{ba})
\sqrt{\frac{A_{ba}^2}{B_{ba}^2}}
\Bigg(
2B_{ba,\alpha}\mathcal P_{b,\beta\nu}
&\\
-B_{ba,\nu}\mathcal P_{b,\alpha\beta}
\Bigg)
+P_{n+1}'(v_{ba})
\Bigg(
A_{ba,\nu}\mathcal P_{b,\alpha\beta}
-2A_{ba,\alpha}\mathcal P_{b,\beta\nu}
\Bigg)
\Bigg]p_a^\alpha p_a^\beta
\end{aligned}
&\\[2mm]
&=
\begin{aligned}[t]
-\frac12 \frac{1}{\sigma_a^2}
\sum_{n=0}^{\infty}
\left(\frac{\ln\sigma_a}{\sigma_a}\right)^n
\sum_{b\neq a}^N
\frac{4G}{m_a^2m_b(A_{ba}^2)^{3/2}}
\left(\frac{B_{ba}^2}{A_{ba}^2}\right)^{n/2}
\Bigg[
P_n'(v_{ba})
\sqrt{\frac{A_{ba}^2}{B_{ba}^2}}
\Delta C_{ba,\beta}
\Bigg(m_b^2p_a^\mu p_a^\beta
&\\
+(p_a\cdot p_b)\left(p_a^\mu p_b^\beta
+2p_b^\mu p_a^\beta\right)
+\left(\frac{(p_a\cdot p_b)^2}{m_b^2}
+\frac{m_a^2}{2}\right)p_b^\mu p_b^\beta
-\eta^{\mu\beta}
\left[
(p_a\cdot p_b)^2-\frac{m_a^2m_b^2}{2}
\right]
\Bigg)
&\\
+P_{n+1}'(v_{ba})
\Bigg(
\frac{m_b^2m_a}{2}p_a^\mu
+\left[
\frac{3m_a(p_a\cdot p_b)}{2}
-\frac{(p_a\cdot p_b)^3}{m_b^2m_a}
\right]p_b^\mu
\Bigg)
\Bigg].&
\end{aligned}
\end{align*}
Comparing this expression with~\eqref{eq:trajectory_acceleration} we identify the leading logarithmic correction:
\begin{equation}
  \label{eq:result_C0}
  C_{0}^{(a)\mu}
  =\sum_{b\neq a}^N{2G\over m_b m_a^2(A_{ba}^2)^{3/2}}
  \left[
    {m_b^2m_a\over2}p_a^\mu
    +\left({3m_a(p_b\cdot p_a)\over2}
    -{(p_b\cdot p_a)^3\over m_b^2 m_a}\right)p_b^\mu
  \right],
\end{equation}
and every (sub)$^n$-leading correction to $\xi_a$:
\begin{align}
  \label{eq:exp_for_Cn_simplified}
  \nonumber
  C_n^{(a)\mu}
  ={}\frac{2G}{n(n+1)}\sum_{b\neq a}
  \frac{1}{m_a^2m_b}\frac{1}{(A^2_{ba})^{3/2}}
  \left(\frac{B_{ba}^2}{A^2_{ba}}\right)^{n/2}\Bigg[P'_n(v_{ba})\sqrt{\frac{A_{ba}^2}{B_{ba}^2}}\Delta C_{ba,\beta}\Bigg(m_b^2p_a^\mu p_a^\beta
  &\\
  \nonumber
  +(p_a\cdot p_b)\left(p_a^\mu p_b^\beta
  +2p_b^\mu p_a^\beta\right)+\left(\frac{(p_a\cdot p_b)^2}{m_b^2}+\frac{m_a^2}{2}\right)p_b^\mu p_b^\beta-\eta^{\mu\beta}\left[(p_a\cdot p_b)^2-\frac{m_a^2m_b^2}{2}\right]\Bigg)
  &\\
  +P_{n+1}'(v_{ba})\Bigg(\frac{m_b^2m_a}{2}p_a^\mu+\left[\frac{3m_a(p_a\cdot p_b)}{2}-\frac{(p_a\cdot p_b)^3}{m_b^2m_a}\right]p_b^\mu\Bigg)\Bigg],&
\end{align}
which is of order $G^{n+1}.$

\subsection{Comparison with known results}

The papers~\cite{saha2020proofclassicalsoftgraviton,sahoo2020classicalsubsubleadingsoftphoton} organize the field equations in a slightly different way. In particular, they multiply both sides of Einstein's equation by $\sqrt{-g}$ in such a way that their matter energy momentum tensor is given solely by
\begin{equation}
  \label{eq:SS_energy_momentum_Tensor}
  T^{\xi,\mu\nu} = \sum_{a=1}^{N} m_a \int_0^\infty \dot\xi^\mu_a(\sigma_a)\dot\xi^\nu_a(\sigma_a) \delta^{(4)}(x-\xi_a(\sigma_a)) d\sigma_a.
\end{equation}

In both papers the result is presented in momentum space for $\hat T^{\xi,\mu\nu}(k)$, but we avoid taking the Fourier transform by using equation 2.11 from~\cite{saha2020proofclassicalsoftgraviton} which can be rewritten as
\begin{equation}
    \int \Box^{-1}T^{\xi,\mu\nu}(x) e^{i\omega t}\, dt = -\frac{1}{4\pi r}e^{i\omega r}\hat T^{\xi,\mu\nu}(k)+\mathcal O(r^{-2}).
\end{equation}
Taking the inverse Fourier transform with respect to time gives
\begin{equation}
    \Box^{-1}T^{\xi,\mu\nu} = -\int\frac{1}{4\pi r}\hat T^{\xi,\mu\nu}(k)e^{-i\omega u}\, \frac{d\omega}{2\pi}+\mathcal O(r^{-2}),
\end{equation}
where $k^\mu=\omega q^\mu$ and again $u=t-r$. The calculation of $\Box^{-1}T^{\xi,\mu\nu}$ now closely follows the derivation of the electromagnetic potential $A^{\mu}$ in \cite{karan2025orderclassicalelectromagneticsoft}, with an extra trajectory factor. The final result is
\begin{align}
  \nonumber
    \Box^{-1}T^{\xi,\mu\nu}=\frac{1}{4\pi r}\sum_{a=1}^N\Bigg[\frac{p^\mu_a p^\nu_a}{q\cdot p_a} + \sum_{n=0}^{\infty}\frac{(\ln u)^{n}}{u^{n+1}}(-1)^{1+n}\Bigg\{\Bigg(
    p^\mu_a C_0^{(a)\nu}+p^\nu_a C_0^{(a)\mu}-p^\mu_a p^\nu_a \frac{q\cdot C_0^{(a)}}{q\cdot p_a}
    \Bigg)
    &\\
    \times
    \left(q\cdot C_0^{(a)}\right)^n
    \label{eq:full_expression_t_mu_nu}
    +\sum_{j=1}^n j\left(\frac{q\cdot p_a}{m_a}\right)^j\Bigg(
      p^\mu_a C_j^{(a)\nu}+p^\nu_a C_j^{(a)\mu}-p^\mu_a p^\nu_a \frac{q\cdot C_j^{(a)}}{q\cdot p_a}
      \Bigg)\binom{n}{j}\left(q\cdot C_0^{(a)}\right)^{n-j}\Bigg\}\Bigg].
\end{align}
Substituting $n=0$ reproduces eq. (3.40) in~\cite{saha2020proofclassicalsoftgraviton}. For $n=1$,~\eqref{eq:exp_for_Cn_simplified} gives
\begin{align}
  \nonumber
  \label{eq:exp_for_C1}
  C_1^{(a)\mu}
  =G\sum_{b\neq a}\frac{m_a m_b^2\Delta C_{ba,\beta}}
  {((p_a\cdot p_b)^2-m_a^2m_b^2)^{3/2}}\Bigg\{\Bigg(\frac{(p_a\cdot p_b)^2}{m_b^2}p_b^\mu p_b^\beta
  +
  (p_a\cdot p_b)p_a^\mu p_b^\beta
  +2(p_a\cdot p_b)p_b^\mu p_a^\beta
  &\\
  \nonumber
  +m_b^2p_a^\mu p_a^\beta+\frac{m_a^2}{2}p_b^\mu p_b^\beta
  -\eta^{\mu\beta}\left[(p_a\cdot p_b)^2-\frac{m_a^2m_b^2}{2}\right]\Bigg)+3{(p_{a}\cdot p_{b})p_b^\beta+m_b^2p_a^\beta\over (p_a\cdot p_b)^2-m_a^2m_b^2}
  &\\
  \times\Bigg(\frac{m_b^2m_a^2}{2}p_a^\mu
  +\left[\frac{3m_a^2(p_a\cdot p_b)}{2}
  -\frac{(p_a\cdot p_b)^3}{m_b^2}\right]p_b^\mu\Bigg)\Bigg\},&
\end{align}
and by plugging~\eqref{eq:result_C0} and~\eqref{eq:exp_for_C1} into~\eqref{eq:full_expression_t_mu_nu}
we reproduce the sub-subleading terms at order $G^2$ from appendix B, Eqs. (B.9), (B.11), and (B.12) of \cite{sahoo2020classicalsubsubleadingsoftphoton}. In both cases, the coefficients of the $u$ expansion differ from the coefficients in the $\omega$ expansion by a factor of $-1$ and $-2i$, respectively, due to the Fourier transform. See appendix C of that same paper.

\section{Discussion and conclusion}

The main result of this work is that, under the assumption that there are no terms like~\eqref{eq-intro:forbiden-soft-log-form-retarded-time}
in the expression for the $r^{-1}$ radiative field near null infinity, the leading order trajectory coefficients at $\lambda_\sigma^0$ may be calculated using only the matter energy-momentum tensor, reducing the problem to a linear one. The expressions for $C_0$ and $C_1$ reproduce the expected logarithmic correction to the asymptotic trajectories. While the agreement with $C_0$ was expected, as it is of order $G$, the agreement with $C_1$ shows that even at $\mathcal O(G^2)$ the coefficients may be calculated solely with the matter energy-momentum tensor. The present work may also serve as a check of the hypothesis that the radiative sector of the $1/r$ field at null infinity is at most $\lambda_U^{-1}$: if under the hypothesis of the problem, the leading order contribution of the relevant part of a soft-log trajectory coefficient is not given by~\eqref{eq:exp_for_Cn_simplified}, then it hints that there must exist relevant terms of the form~\eqref{eq-intro:forbiden-soft-log-form-retarded-time}.

The simplicity of the $\lambda_\tau^{-1}$ sector of the metric also raises the question of whether one can produce a similar analysis for the $\lambda^{-2}_\tau$ sector of $Q^{\mu\nu}$, since this sector should directly contribute to soft log terms at null infinity. At this order one cannot ignore the non-linear source, since it directly enters the equation for $Q^{2,\mu\nu}_n$. However, the block separation introduced in section~\ref{sec:lambda_field} may offer a clearer view of which terms should contribute. Moreover, we point out that the homogeneous solution associated with the $\lambda_\tau^{-2}$ sector in~\eqref{eq:complete_Pi_block} gives
\begin{equation*}
  \int A^2_{n-1}(\hat q)\frac{(\ln(\tau \kappa))^{n-1}}{(\tau \kappa)^{n+1}}d^2\hat q\rightarrow \frac{2\pi}{n}A^2_{n-1}(\hat x)\frac{(\ln u)^{n-1}}{ru^n}
\end{equation*}
at null infinity. These terms have precisely the form of the soft log terms. Since these homogeneous modes represent what we lose after using only $\mathcal H_n^{-1}$ to invert the non-linear source, it is natural to ask whether properly inverting only the $\lambda_\tau^{-4}$ sector of the non-linear source using the Green function provides all the information needed to reconstruct the soft-sector. More specifically, we ask whether the leading soft-sector can be completely characterized by
\begin{equation}
  \Box^{-1}\left(16\pi G(-g)T^{\mu\nu} + N^{\mu\nu}\left[\mathfrak h_M|_{\lambda^{-1}_\tau},\mathfrak h_M|_{\lambda^{-1}_\tau}\right]\right)\Big|_{r^{-1},\lambda_u^{-1}},
\end{equation}
and we showed that the results of this work allow one to calculate the contribution from the energy momentum tensor in exactly the same way as in~\cite{karan2025orderclassicalelectromagneticsoft}. We leave this question for future work.

\acknowledgments

I would like to thank Alberto Saa for useful discussions and feedback on this draft. This study was financed in part by the Coordenação de
Aperfeiçoamento de Pessoal de Nível Superior - Brasil (CAPES) -
Finance Code 001.

\nocite{*}
\bibliographystyle{JHEP}
\bibliography{refs}
\end{document}